%% file: modal-method.tex
\documentclass[prb,amsmath,twocolumn,floatfix]{revtex4-2}

\usepackage{amsfonts}
\usepackage{graphicx}
\usepackage{mathtools}
\usepackage{textcomp}
\usepackage{array}

\usepackage[T1]{fontenc}
\usepackage[utf8]{inputenc}
\usepackage{lmodern}

\renewcommand*{\arraystretch}{1.5}

\DeclareMathSymbol{\Phi}{\mathalpha}{letters}{8}

\newcommand\I{\mathrm{i}}
\newcommand\E{\mathrm{e}}

\newcommand\tsub[1]{_{\text{#1}}}

\newcommand\vect[1]{\boldsymbol{#1}}
\newcommand\mat[1]{\mathsf{#1}}
\newcommand\inner[2]{\bigl\langle #1, #2 \bigr\rangle}
\newcommand\bigginner[2]{\biggl\langle #1, #2 \biggr\rangle}
\newcommand\abs[1]{\lvert #1 \rvert}
\newcommand\px{\partial_x}
\newcommand\pz{\partial_z}
\newcommand\diff{\mathrm{d}}
\newcommand\transposed{^{\mathrm{T}}}
\newcommand\smallfactor[1]{\underline{#1}}
\DeclareMathOperator{\RE}{Re}
\DeclareMathOperator{\IM}{Im}

\begin{document}

\title{A modal approach to modelling spin wave scattering}

\author{Wojciech Śmigaj}
\email{w.smigaj@optopol.com.pl}
\affiliation{Met Office, FitzRoy Rd, Exeter, EX1 3PB, UK}
\altaffiliation{Present address: Optopol Technology, ul. Żabia 42, 42-400 Zawiercie, Poland}

\author{Krzysztof Sobucki}
\email{krzsob@amu.edu.pl}
\affiliation{Institute of Spintronics and Quantum Information, Faculty of Physics, Adam Mickiewicz University, Uniwersytetu Poznańskiego 2, 61-614 Poznań, Poland}

\author{Paweł Gruszecki}
\email{gruszecki@amu.edu.pl}
\affiliation{Institute of Spintronics and Quantum Information, Faculty of Physics, Adam Mickiewicz University, Uniwersytetu Poznańskiego 2, 61-614 Poznań, Poland}

\author{Maciej Krawczyk}
\email{krawczyk@amu.edu.pl}
\affiliation{Institute of Spintronics and Quantum Information, Faculty of Physics, Adam Mickiewicz University, Uniwersytetu Poznańskiego 2, 61-614 Poznań, Poland}

\begin{abstract}
  Efficient numerical methods are required for the design of optimised devices. In magnonics, the primary computational tool is micromagnetic simulations, which solve the Landau-Lifshitz equation discretised in time and space. However, their computational cost is high, and the complexity of their output hinders insight into the physics of the simulated system, especially in the case of multimode propagating wave-based devices. We propose a finite-element modal method allowing an efficient solution of the scattering problem for dipole-exchange spin waves propagating perpendicularly to the magnetisation direction. The method gives direct access to the scattering matrix of the whole system and its components. We extend the formula for the power carried by a magnetostatic mode in the Damon-Eshbach configuration to the case with exchange, allowing the scattering coefficients to be normalised to represent the fraction of the input power transferred to each output channel. We apply the method to the analysis of spin-wave scattering on a basic functional block of magnonic circuits, consisting of a resonator dynamically coupled to a thin film. The results and the method are validated by comparison with micromagnetic simulations.
\end{abstract}

\maketitle

\section{Introduction}

In recent years, we have observed a rapid progress in the development of components for magnonic circuitry. Conduits for single and multimode spin-wave transfer \cite{Sadovnikov2015,Rana2018,Bjorn2021,Haldar2021,Sahoo2021}, phase control \cite{Louis2016,Wang2018-2,Dobrovolskiy2019,Baumgaertl2021}, spin-wave valves \cite{Au2021}, couplers \cite{Wang2018,Graczyk2018}, resonators \cite{Qin2021,Sobucki2021,Grachev2021}, transducers, diodes \cite{Szulc2020}, and logic gates \cite{Kostylev2005,Fischer2017} are only selected examples  based on various physical principles. To understand the physics of the spin-wave phenomena behind the observed functionalities,  increase the effectiveness of their operation and find their new realisations, researchers need suitable models and numerical methods. 

The primary approach used in magnonics is micromagnetism, where the nonlinear Landau-Lifshitz (LL)  torque equation is used to describe magnetisation dynamics. It is usually solved in time and space with micromagnetic simulations based on the continuum model \cite{Kumar_2017,Abert2019}. These methods offer a faithful description of the experimental realisations, including nonlinear and temperature effects. There are two principal implementations of micromagnetic solvers, one based on the finite difference \cite{vansteenkiste2014design,Lepadau2020} and the second on the finite element method \cite{Schrefl2007}. However, micromagnetic simulations are time-consuming and require extensive computational power. Their outputs are raw time- and space-dependent data, and extensive postprocessing  is necessary to elucidate the physical mechanisms underlying complex magnetic systems. In addition, simulations of wave dynamics over time require selecting a source of these waves; the obtained spectrum is source-dependent and mode identification may be ambiguous.

Other approaches are based on solving the LL equation in the frequency domain and either wavevector or real space; they are commonly referred to as spectral methods. Spectral methods enable calculation of the response of a magnetic system to a time-harmonic excitation with high precision and at a lower computational cost, though at the price of approximations, one of which is linearisation. An example of a spectral method is the plane wave method, applicable to systems with discrete translational symmetry. It was introduced and used to calculate the band structure of bulk~\cite{Krawczyk2008} and thin-film magnonic crystals \cite{Sokolovskyy2011,Gallardo2018,Chang2018} as well as magnonic quasicrystals \cite{Rychly2018,Watanabe2021}. In the latter case, a full magnetic saturation and homogeneity across the film thickness were assumed. Here, the LL equation is transformed into an infinite set of algebraic equations in the frequency and wavevector domain. The equations are indexed by the reciprocal lattice vectors. The eigenproblem formed by the truncated system is solved numerically with standard numerical routines. 

The dynamical matrix method~\cite{Grimsditch2004,Tacchi2011} overcomes some limitations of the plane wave method. It uses the finite-difference method to solve the LL equation formulated in real space and linearised about the magnetisation ground state derived from micromagnetic simulations. This is a powerful method used to calculate normal modes in isolated nanoelements~\cite{Grimsditch2004} and thin-film magnonic crystals with nanoscale periodicity \cite{Tacchi2011}. But the large matrices involved here make the solution of the eigenproblem and analysis of the normal modes time-consuming. Furthermore, the method does not make it possible to study spin-wave scattering and transmission.
An extension of the normal mode calculation method proposed in Ref.~\onlinecite{Daquino2009} allows damping and, to some extent, nonlinear effects to be taken into account. The resulting eigenproblem can be discretised either with finite differences or finite elements \cite{perna2021}. However, these methods do not encompass calculation of the transmission, reflection, or scattering matrices for spin waves in nanoscale objects.

\begin{figure}
 \centerline{\includegraphics{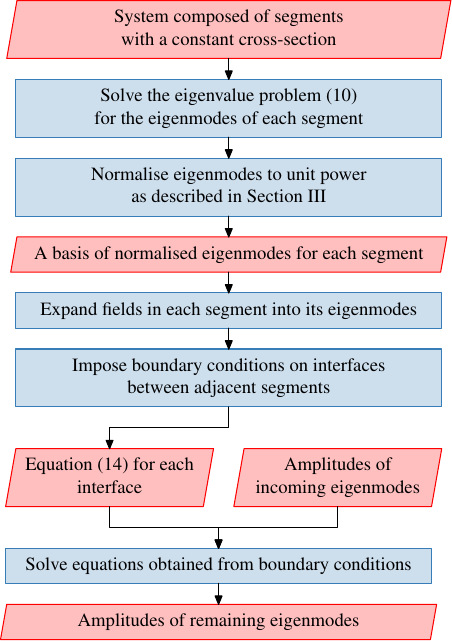}}
 \caption{Steps required for the solution of a scattering problem with the proposed modal method. Inputs and outputs are denoted with parallelograms and processes with rectangles.\label{fig:flowchart}}
\end{figure}

In this paper, we develop an efficient finite-element modal method to solve the scattering problem for dipole-exchange spin waves propagating in a system composed of one or more ferromagnetic layers or stripes magnetised perpendicularly to the wave propagation direction. The proposed computational procedure is outlined in Fig.~\ref{fig:flowchart}. We decompose the system into segments with a constant cross-section and find the normal modes of each segment by solving the linearised LL equation discretised with finite elements. We expand the dynamical components of the magnetisation and magnetostatic potential in each segment in the basis of its normal modes. These expansions are tied together by imposing appropriate boundary conditions on each interface. The resulting system of linear equations is solved for the amplitudes of outgoing (transmitted and reflected) modes produced by incoming modes with known amplitudes. Optionally, the subset of these equations associated with a particular interface can also be solved for the scattering matrix of that interface. Interfacial scattering matrices supply valuable information about the contribution of individual scattering pathways to the output signal and the role of particular normal modes excited within each segment. This yields a deeper insight into the physics of the system under consideration and can help its designer optimise its geometry for a specific application.

Another contribution of this paper is the generalisation of the Lorentz reciprocity theorem, mode orthogonality relations, and the formula for the power carried by propagating spin-wave modes of a tangentially magnetised multilayer to the case of dipole-exchange waves. These results enable propagative normal modes used in field expansions to be normalised to unit power, letting squared scattering coefficients be identified with the power passed to the corresponding scattering channels.

We use the proposed method to study the transmission and reflection of spin waves on a ferromagnetic stripe coupled with a ferromagnetic film. This system can be considered as a basic building block of magnonic circuits possessing various functionalities \cite{Au2012,Yu2019,Yu2019b,Szulc2020,Qin2021,Sobucki2021,Wang2021,Pierre2021,Fripp2021}.  We elucidate the important role played by pairs of modes with contrasting group velocities supported by the bilayer formed by the film and the stripe. The validity of the modal method is confirmed by an excellent agreement of its predictions with results of micromagnetic simulations.

The paper is organised as follows. In the next section, we describe the finite-element modal method, discussing first the determination of eigenmodes (Sec.~\ref{sec:determination-of-eigenmodes}) and then the mode-matching equations (Sec.~\ref{sec:mode-matching}). In Sec.~\ref{sec:power-flow} and Appendix~\ref{sec:ortogonality-relations} we derive the Lorentz reciprocity theorem, mode orthogonality relations, a formula for the power carried by dipole-exchange spin waves in the Damon-Eshbach configuration. In Sec.~\ref{sec:applications} we use the proposed method to analyse the scattering of spin waves propagating along a thin ferromagnetic film on a resonant element placed in its vicinity. After validating its predictions against results of micromagnetic simulations, we feed the calculated mode propagation constants and scattering coefficients into a semi-analytical model of the system under consideration, which allows us to explain the physical origin of notable features visible in its reflection and transmission spectra.

\section{Finite-element modal method}

\subsection{Introduction}
\label{sec:femm-introduction}

\begin{figure}
 \centerline{\includegraphics{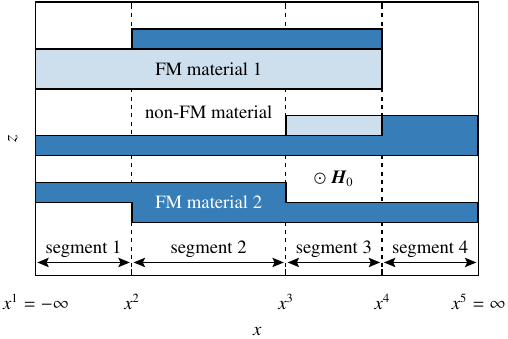}}
 \caption{An example system whose geometry satisfies the assumptions made in Sec.~\ref{sec:femm-introduction}. The system contains four segments with uniform cross-sections in the $xy$~plane. FM = ferromagnetic.\label{fig:generic-system}} 
\end{figure}

We consider a system composed of ferromagnetic and non-ferromagnetic materials. Its geometry is independent of the $y$~coordinate and piecewise constant along $x$. The system is placed in an external static magnetic field oriented along the $y$~axis; this field is assumed to be sufficiently strong to saturate all magnetic materials and orient their static magnetisation along~$y$. An example of such a system is shown schematically in Fig.~\ref{fig:generic-system}. 

In a modal method, each $x$-invariant region is treated as a finite or semi-infinite segment of a waveguide with a uniform cross-section. The fields inside each segment are expressed as a superposition of eigenmodes of the corresponding waveguide. These eigenmode expansions are coupled through boundary conditions imposed on the interfaces $x = x^i$ ($i = 2, 3, \dots, n$, where $n$~is the number of segments) separating adjacent segments; imposition of these conditions produces a linear system of equations for the mode amplitudes (excitation coefficients). Typically, the excitation coefficients of modes incoming from the left and right are known, and the quantities of interest, obtained by solving the system of equations, are the coefficients of the outgoing modes in the first and last segment.

\subsection{Determination of waveguide eigenmodes}
\label{sec:determination-of-eigenmodes}

As said above, the fields in each $x$-invariant segment of the system are expanded in the eigenmodes, both propagative and evanescent, of an $x$- and $y$-invariant waveguide whose profile along~$z$ matches that of the segment. These eigenmodes are determined numerically using the finite-element method. The waveguide is governed by the Gauss law for magnetism (applicable everywhere):
\begin{equation}
 \vect\nabla \cdot \vect B = 0
\end{equation}
and the LL equation with a Gilbert damping term \cite[Sec.~3.8]{stancil2009spin} (applicable only in the ferromagnetic layers):
\begin{equation}
 \partial_t{\vect M} = \gamma \mu_0 \vect M \times \vect H\tsub{eff} + \frac{\alpha}{M\tsub S} \vect M \times \partial_t \vect M,
\end{equation}
where $\vect M$ is the magnetisation, $\vect H\tsub{eff}$ the effective magnetic field, $\gamma$ the gyromagnetic ratio, $\mu_0$ the vacuum permeability, $M\tsub S$ the saturation magnetisation and $\alpha$ the damping coefficient. The effective magnetic field is taken to be a superposition of the static external magnetic field~$\vect H_0$, the magnetostatic magnetic field~$\vect H\tsub{m}$ and the exchange magnetic field~$H\tsub{ex}$:
\begin{equation}
 \vect H\tsub{eff} = \vect{H}_0 + \vect H\tsub{m} + \vect H\tsub{ex}.
\end{equation}
Assuming a harmonic time dependence [$\exp(-\I \omega t$)], splitting the magnetisation $\vect M$ and magnetostatic magnetic field $\vect H\tsub m$ into static and dynamic (radio-frequency) components, expressing the latter as a gradient of the magnetostatic potential ($\vect h = -\nabla \phi$), writing the exchange magnetic field as $\vect H\tsub{ex} = \vect\nabla \cdot (l^2 \vect\nabla \vect m)$ (with $l$ denoting the exchange length) \cite{Krawczyk2012} and linearising the LL equation, we arrive at the following system of equations:
\begin{subequations}
\label{eq:linearised-equations}
\begin{align}
  \partial_x(m_x - \partial_x \phi) + \partial_z(m_z - \partial_z \phi) &= 0, \\  
  \partial_x \phi - \vect\nabla \cdot (l^2 \vect\nabla m_x) + 
  \frac{H_0}{M_S} m_x + \frac{\I \omega}{\gamma \mu_0 M_S} (m_z + \alpha m_x) &= 0, \\
  \partial_z \phi - \vect\nabla \cdot (l^2 \vect\nabla m_z) + 
  \frac{H_0}{M_S} m_z - \frac{\I \omega}{\gamma \mu_0 M_S} (m_x - \alpha m_z) &= 0,
\end{align}
\end{subequations}
where all material coefficients are functions of $z$ only. The $\phi$, $m_x$ and $m_z$ fields of a waveguide eigenmode have a harmonic dependence on $x$:
\begin{equation}
 \begin{Bmatrix}
  \phi(x, z)\\
  m_x(x, z)\\
  m_z(x, z)
 \end{Bmatrix}
 =
 \begin{Bmatrix}
  \phi(z)\\
  m_x(z)\\
  m_z(z)
 \end{Bmatrix}
 \exp(\I k_x x),
\end{equation}
where $k_x$ is the mode wavenumber. Taking advantage of this fact and introducing the symbols $\tilde m_x \coloneqq \I m_x$, $\omega_M \coloneqq -\gamma \mu_0 M_S$ and $\omega_0 \coloneqq -\gamma \mu_0 H_0$, we can rewrite the equations in the form
\begin{subequations}
\label{eq:governing-eqs-wg-mode}
\begin{align}
  k_x(\tilde m_x + k_x\phi) + \partial_z(m_z - \partial_z \phi) &= 0, \\  
  k_x \phi + \partial_z (l^2 \partial_z \tilde m_x) - k_x^2 l^2 \tilde m_x -
  \frac{\omega_0 - \I\alpha\omega}{\omega_M} \tilde m_x - \frac{\omega}{\omega_M} m_z &= 0, \\
  \partial_z \phi - \partial_z (l^2 \partial_z m_z) + k_x^2 l^2 m_z +
  \frac{\omega_0 - \I\alpha\omega}{\omega_M} m_z + \frac{\omega}{\omega_M} \tilde m_x &= 0.
\end{align}
\end{subequations}

This is a quadratic eigenvalue problem in $k_x$; its approximate solution can be found by discretising the above equations with the finite-element method. This requires transforming them into a \emph{weak form}, which forces the integrals of their residuals weighted with an appropriate set of \emph{test functions} to vanish \cite{Hughes2000,Jin2015}. To this end, we multiply the equations by test functions $\psi$, $\tilde n_x$ and $n_z$, respectively, and integrate by parts over $z$ to reduce the order of differentiation and thus lower the smoothness requirements on the \emph{trial functions} into which $\phi$, $\tilde m_x$ and $m_z$ will be expanded. Application of the boundary and continuity conditions
\begin{itemize}
 \item $\lim_{z\to\pm\infty}\phi(z) = 0$,
 \item $b_z \equiv m_z - \pz\phi$ is continuous along $z$,
 \item $l^2 \pz m_x$ and $l^2 \pz m_z$ are continuous along $z$ (which implies in particular that $\pz m_x = \pz m_z = 0$ on interfaces between layers with and without exchange magnetic field)
\end{itemize}
annihilates the boundary terms produced by integration by parts and leads to the following weak form: Find $k_x$, $\phi$, $\tilde m_x$ and $m_z$ such that for all $\psi$, $\tilde n_x$ and $n_z$
\begin{widetext}
\begin{subequations}
\label{eq:weak-form}
\begin{align}
  \inner{\pz \psi}{\pz \phi} + k_x^2 \inner{\psi}{\phi} 
  + k_x \inner{\psi}{\tilde m_x} - \inner{\pz \psi}{m_z} &= 0,\\  
  k_x \inner{\tilde n_x}{\phi} 
  - \inner{\pz \tilde n_x}{l^2 \pz \tilde m_x} 
  - k_x^2 \inner{\tilde n_x}{l^2 \tilde m_x} 
  - \bigginner{\tilde n_x}{\frac{\omega_0 - \I\alpha\omega}{\omega_M}\tilde m_x}
  - \bigginner{\tilde n_x}{\frac{\omega}{\omega_M}m_z}
  &= 0,\\
  - \inner{n_z}{\pz \phi} 
  - \bigginner{n_z}{\frac{\omega}{\omega_M}\tilde m_x}
  - \inner{\pz n_z}{l^2 \pz m_z} - k_x^2 \inner{n_z}{l^2 m_z} 
  - \bigginner{n_z}{\frac{\omega_0 - \I\alpha\omega}{\omega_M}m_z} 
  &= 0,
\end{align}
\end{subequations}
\end{widetext}
where
\begin{equation}
 \inner{f}{g} \coloneqq \int f(z)\,g(z)\,\mathrm{d}z.
\end{equation}
The solutions $(\phi, \tilde m_x, m_z)$ and the test functions $(\psi, \tilde n_x, n_z)$ are required to satisfy the \emph{essential} boundary and continuity conditions, i.e.\ those not involving derivatives.

This weak form can be discretised using the Galerkin method. The fields $\phi$, $\tilde m_x$ and $m_z$ are expressed as finite linear combinations of appropriate basis functions (defined on a sufficiently long but finite interval $z\tsub{min} \leq z \leq z\tsub{max}$ in the case of $\phi$ and on the union of the intervals where $l^2(z) > 0$ in the cases of $\tilde m_x$ and $m_z$), and the weak form is evaluated with the test functions $\psi$, $\tilde n_x$ and $n_z$ set to each of these basis functions in turn. This leads to a quadratic algebraic eigenvalue problem
\begin{equation}
 {\mat A \vect x} + k_x {\mat B \vect x} + k_x^2 {\mat C \vect x} = 0,
\end{equation}
where $\vect x$ is the vector of expansion coefficients of $\phi$, $\tilde m_x$ and $m_z$, and $\mat A$, $\mat B$ and $\mat C$ are matrices independent from $k_x$. This quadratic eigenvalue problem can be rewritten as a generalised linear eigenvalue problem:
\begin{equation}
 \label{eq:gep}
 \begin{bmatrix}
  \mat A & \mat B\\ & \mat I
 \end{bmatrix}
 \begin{bmatrix}
  \vect x \\ \vect y
 \end{bmatrix}
  = k_x
 \begin{bmatrix}
  & -\mat C \\ \mat I &
 \end{bmatrix}
 \begin{bmatrix}
  \vect x \\ \vect y
 \end{bmatrix}
 ,
\end{equation}
which can be solved using the standard QZ algorithm~\footnote{In practice, for better numerical stability, it is advantageous to rewrite the weak form in terms of the scaled variables $\tilde m_x' \coloneqq a \tilde m_x$, $m_z' \coloneqq a m_z$ and $k_x' \coloneqq a k_x$, where $a$ is a length comparable with the thickness of the ferromagnetic layers. This ensures the elements of all matrix blocks have similar magnitude.}. When damping is neglected, all matrices in the eigenproblem written in terms of $\phi$, $\tilde m_x$ and $m_z$ are real, so the phases of eigenvectors corresponding to propagative modes (modes with real $k_x$) can be chosen so that the profiles $\phi(z)$ and $m_z(z)$ are real whereas $m_x(z)$ is imaginary.

\subsection{Mode matching}
\label{sec:mode-matching}

The fields in $i$th $x$-invariant waveguide segment, sandwiched between the planes $x=x^i$ and $x=x^{i+1}$, are expanded in the basis of eigenmodes determined as described in the previous section:
\begin{multline}
 \phi(x, z) = \sum_j U^i_j(x)\, \phi^{iu}_j(z) +
              \sum_j D^i_j(x)\, \phi^{id}_j(z) \\
 \text{for} \quad x^i \leq x \leq x^{i+1},
\end{multline}
and similarly for $m_x$ and $m_z$. The first sum runs over modes propagating or decaying rightwards (towards $x= \infty$); the second, over modes propagating or decaying leftwards (towards $x=-\infty$). The symbol $U^i_j(x)$ denotes the position-dependent excitation coefficient of $j$th rightward mode of $i$th segment with magnetostatic potential profile $\phi^{iu}_j(z)$. Analogous symbols containing the letter~$d$ are used for leftward modes. Within each segment, $U^i_j(x)$ and $D^i_j(x)$ vary harmonically and can be written as
\begin{subequations}
\begin{align}
 U^i_j(x) &= U^i_j(x^{iu}) \exp[\I k^{iu}_{xj} (x - x^{iu})],\\
 D^i_j(x) &= D^i_j(x^{id}) \exp[\I k^{id}_{xj} (x - x^{id})].
\end{align}
\end{subequations}
where $k^{iu}_{xj}$ and $k^{id}_{xj}$ are mode wavenumbers. It is convenient to choose the reference positions $x^{iu}$ and $x^{id}$ as
\begin{equation}
 x^{iu} = x^{\max(i, 2)}, \quad
 x^{id} = x^{\min(i+1, n)}.
\end{equation}
This ensures that imposition of boundary conditions on segment interfaces leads to equations [Eq.~\eqref{eq:mode-match} below] containing exponentials whose magnitude does not exceed 1, which could compromise numerical stability.

The fields in adjacent waveguide segments are linked by the following boundary conditions that must hold on the interfaces between these segments:
\begin{itemize}
 \item $\phi$ is continuous along~$x$ on the whole interface,
 \item $b_x$ is continuous along~$x$ on the whole interface,
 \item $m_x$ and $m_z$ are continuous along~$x$ on interfaces separating pairs of layers such that $l^2 > 0$ in both layers,
 \item $l^2 \partial_x m_x$ and $l^2 \partial_x m_z$ are continuous along~$x$ on interfaces separating pairs of layers such that $l^2 > 0$ in at least one layer.
\end{itemize}
These boundary conditions are imposed by multiplying them with the basis functions used to expand the fields in all layers (for the first two boundary conditions) or in the layers fulfilling the specified criteria (for the last two boundary conditions) and integrating over $z$. For each interface, this leads to a set of linear equations that can be written symbolically as
\begin{multline}
 \label{eq:mode-match}
 \begin{bmatrix}
  \mat V^{iu} & \mat V^{id}
 \end{bmatrix}
 \begin{bmatrix}
  \mat E^{iu}(x^{i+1} - x^{iu}) \\
  & \mat I
 \end{bmatrix}
 \begin{bmatrix}
  \vect u^i \\
  \vect d^i
 \end{bmatrix}\\
 =
 \begin{bmatrix}
  \mat W^{i+1,u} & \mat W^{i+1,d}
 \end{bmatrix}
 \begin{bmatrix}
  \mat I \\
  & \mat E^{i+1,d}(x^{i+1} - x^{i+1,d})
 \end{bmatrix}
 \begin{bmatrix}
  \vect u^{i+1} \\
  \vect d^{i+1}
 \end{bmatrix}
 ,
\end{multline}
where $i+1$ is the index of the interface. In this formula, $\mat E^{\cdots}(\Delta x)$ are diagonal matrices of exponentials dependent on the length of segments adjacent to $x^{i+1}$, whereas the matrices $\mat V^{\cdots}$ and $\mat W^{\cdots}$ are independent from that length. Explicit expressions for these matrices are provided in Appendix~\ref{sec:mode-matching-matrices}. The symbols $\vect u^i$ and $\vect d^i$ denote vectors of the excitation coefficients $u^i_j \coloneqq U^i_j(x^{iu})$ and $d^i_j \coloneqq D^i_j(x^{id})$ ($j = 1, 2, \dots$) at the reference positions $x^{iu}$ and $x^{id}$. 

\subsection{Solution of the scattering problem}
\label{sec:solution-of-scattering-problem}

The excitation coefficients of the outgoing modes of the semi-infinite waveguide seg\-ments---as well as the excitation coefficients of modes of any finite segments---can be calculated by solving the system of equations obtained by combining equations of the form~\eqref{eq:mode-match} for $i = 2, 3, \ldots, n$, with the coefficients of the incoming modes, $\vect u^1$ and $\vect d^n$, treated as known. Alternatively, the scattering matrices of individual interfaces can be calculated independently and then concatenated using the algorithm from Ref.~\onlinecite{Li96ScattMat} to reduce the computational expense.

\section{Mode normalisation and power flux}
\label{sec:power-flow}

Frequently, the main quantities of interest in the solution of a scattering problem are the reflectance and transmittance of the structure in question. As we show below, in the absence of damping, these can be identified with the squared magnitudes of the elements of $\vect d^1$ and $\vect u^n$ corresponding to propagating modes, provided that the mode profiles are normalised to unit power. 

Stancil (Ref.~\onlinecite{stancil2009spin}, Sec.~6.1) identifies the time-averaged Poynting vector of exchange-free magnetostatic waves with
\begin{equation}
 \langle \vect S\tsub{exchange-free}\rangle = \tfrac{1}{2} \RE(-\I\omega \phi^* \vect b).
\end{equation}
It follows that, in the exchange-free approximation, the power carried by time-harmonic spin waves propagating along the $x$~axis of an $x$- and $y$-invariant waveguide is given by
\begin{equation}
\label{eq:power-stancil}
\begin{split}
 P\tsub{exchange-free} &=
 \int_{-\infty}^\infty \langle S_{\text{exchange-free},x}\rangle\,\diff z 
 \\
 &= \frac{1}{2} \int_{-\infty}^\infty \RE(-\I\omega \phi^* b_x)\,\diff z 
 \\
 &= \frac{1}{2} \int_{-\infty}^\infty \IM(\omega \phi^* b_x)\,\diff z.
\end{split} 
\end{equation}
We will now generalise this expression to dipole-exchange spin waves (in the Damon-Eshbach configuration). In Appendix~\ref{sec:ortogonality-relations} we derive an orthogonality relation between a pair of dipole-exchange eigenmodes of an $x$- and $y$-invariant waveguide with negligible damping:
\begin{multline*}
 \int_{-\infty}^\infty [\mu_0^{-1} (-\phi_a b_{bx}^* + \phi_b^* b_{ax}) \\
  + \I l^2 (k_{xa} + k_{xb}^*) \vect m_{a} \cdot \vect m_{b}^*]\,\diff z = 0
  \\\text{if}\quad
  k_{xa} \neq k_{xb}^*.
\tag{\ref{eq:lorentz-ortho-same-conjugated}}  
\end{multline*}
Here, $\phi_a(z)$, $\vect m_a(z)$ and $k_{xa}$ are the field profiles and the wavenumber of eigenmode~$a$; $\phi_b(z)$, $\vect m_b(z)$ and $k_{xb}$, those of eigenmode $b$; and $b_{ix}(z) = \mu_0(m_{ix} -\partial_x\phi_i)$ for $i = a, b$. Using the identities $\I k_{xa} \vect m_a = \partial_x \vect m_a$ and $\I k_{xb} \vect m_b = \partial_x \vect m_b$, we can rewrite this relation in a wavenumber-free form:
\begin{multline}
 \label{eq:lorentz-ortho-same-conjugated-no-kx}
 \int_{-\infty}^\infty [\mu_0^{-1} (-\phi_a b_{bx}^* + \phi_b^* b_{ax}) \\
  + l^2(-\vect m_{a} \cdot \partial_x \vect m_{b}^* + \vect m_{b}^* \cdot \partial_x \vect m_{a})]\,\diff z = 0
  \\\text{if}\quad
  k_{xa} \neq k_{xb}^*.
\end{multline}
When $a$ and $b$ refer to the same mode, the integral from the above equation (omitting the now redundant mode index) reduces to:
\begin{equation}
  \begin{split}
 P' &\coloneqq \int_{-\infty}^\infty [\mu_0^{-1} (-\phi b_{x}^* + \phi^* b_{x}) \\
  &\quad\quad\quad\quad+ l^2 (-\vect m \cdot \partial_x \vect m^* + \vect m^* \cdot \partial_x \vect m]\,\diff z \\
  &\hphantom{:}= 2\I \int_{-\infty}^\infty 
  \IM (\mu_0^{-1} \phi^* b_{x} + l^2 \vect m^* \cdot \partial_x \vect m )\,\diff z.
  \end{split}
\end{equation}
Comparison with Eq.~\eqref{eq:power-stancil} shows that 
\begin{equation}
\label{eq:power}
\begin{split}
 P &\coloneqq -\frac14\I \mu_0 \omega P' \\
 &\hphantom{:}= 
  \frac12 \int_{-\infty}^\infty 
  \IM (\omega \phi^* b_{x} + \omega \mu_0 l^2 \vect m^* \cdot \partial_x \vect m )\,\diff z
\end{split}
\end{equation}
reduces to the expression from Eq.~\eqref{eq:power-stancil} when the exchange interaction is neglected, i.e.\ when $l = 0$. This motivates identifying $P$ with the power carried by a dipole-exchange spin wave in the Damon-Eshbach configuration. 

In general, the spin wave will be a superposition of multiple waveguide modes:
\begin{equation}
 \begin{Bmatrix}
  \phi(x, z)\\
  \vect m(x, z)
 \end{Bmatrix}
 =
 \sum_i a_i
 \exp(\I k_{xi} x)
 \begin{Bmatrix}
  \phi_i(z)\\
  \vect m_i(z)
 \end{Bmatrix}
 ,
\end{equation}
where $[\phi_i(z), \vect m_i(z)]$ are the field profiles of the $i$th mode, $k_{xi}$ its wavenumber and $a_i$ its excitation coefficient. From Eq.~\eqref{eq:power}, the total power carried by these modes will be 
\begin{equation}
 \label{eq:power-expansion}
 \begin{split}
  P &= \frac{\I\mu_0\omega}{4} \int_{-\infty}^\infty 
  [\mu_0^{-1} (\phi b_{x}^* - \phi^* b_{x}) \\
  &\qquad\qquad+ l^2 (\vect m \cdot \px\vect m^* - \vect m^* \cdot \px\vect m)]
  \,\diff z \\
  &= \sum_{i,j} a_i a_j^* P_{ij},
 \end{split}
\end{equation}
where
\begin{equation}
 \label{eq:mode-power}
 \begin{split}
  P_{ij} &\coloneqq \frac{\I\mu_0\omega}{4} \int_{-\infty}^\infty 
  [\mu_0^{-1} (\phi_i b_{xj}^* - \phi_j^* b_{xi}) \\
  &\qquad\qquad+ l^2 (\vect m_i \cdot \px\vect m_j^* - \vect m_j^* \cdot \px\vect m_i)]
  \,\diff z.
 \end{split}
\end{equation}
The orthogonality relation \eqref{eq:lorentz-ortho-same-conjugated-no-kx} implies that, in the absence of damping and of degenerate modes, the integral $P_{ij}$ vanishes unless (a) $i = j$ and mode~$i$ is propagative ($k_{xi}$ is real) or (b) mode $i$ is an evanescent mode ($k_{xi}$ is not real) and mode $j$ is its complex-conjugate counterpart ($k_{xj} = k_{xi}^*$). Therefore under these assumptions the total power carried by a superposition of waveguide modes is the sum of powers carried by individual propagative modes and pairs of evanescent modes with complex-conjugate wavenumbers: 
\begin{equation}
 \label{eq:power-of-mode-superposition}
 \begin{split}
  P &= \sum_{\text{propagative modes $i$}} \abs{a_i}^2 P_{ii} \\
  & \quad+ \sum_{\text{evanescent modes $i$}} \RE[a_i a_{\text{conj}(i)}^* P_{i,\text{conj}(i)}],
 \end{split}
\end{equation}
where $\text{conj}(i)$ denotes the index of the mode with wavenumber $k_{xi}^*$. In practice, it is convenient to normalise mode profiles so that $P_{ii} = 1$ for propagative modes and $P_{i, \text{conj}(i)} = 1$ for evanescent ones, since this makes it possible to obtain the power carried by individual modes or pairs of modes directly from their excitation coefficients. 

Importantly, Eq.~\eqref{eq:mode-power} enables unit-power mode normalisation even when the profiles of the incoming and outgoing modes in a system are neither identical nor related by symmetry, e.g.\ when the input and output waveguides are multimodal or have different geometries.

If degenerate modes exist, Eq.~\eqref{eq:power-of-mode-superposition} is still valid provided that such modes have been suitably orthogonalised. On the other hand, when damping is present, Eq.~\eqref{eq:lorentz-ortho-same-conjugated-no-kx} loses its validity, and so the total power cannot in general be decomposed into a simple sum of powers carried by individual modes: the cross terms proportional to $P_{ij}$ with $i \neq j$ do not disappear. However, if the damping is low enough, such a decomposition may still be accurate enough for practical purposes, as will be shown numerically in the next section.

\section{Applications}
\label{sec:applications}

\subsection{Introduction}

In this section we use the finite-element modal method described above to simulate the scattering of spin waves travelling along a thin ferromagnetic film on a stripe of another ferromagnetic material placed above the film. We validate the method by comparing its predictions against results of micromagnetic simulations. Finally, to understand the variation of the scattering coefficients with the stripe width, we develop a semi-analytical model elucidating the roles played by the two pairs of modes supported by the bilayer made of the film and the stripe. The numerical inputs required by the model---mode wavenumbers and scattering matrices---are obtained directly from simulations made with the modal method.

\subsection{The system under consideration \label{subsec:parameters}} 

\begin{figure}
 \centerline{\includegraphics{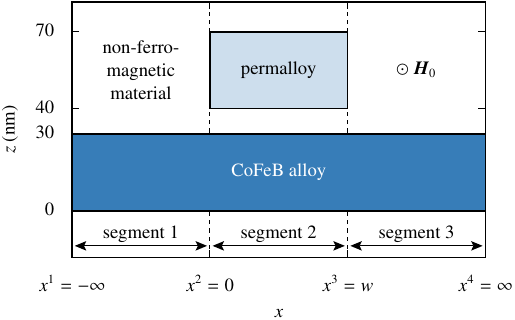}}
 \caption{$xz$-plane cross-section of the $y$-invariant system analysed in Sec.~\ref{sec:applications}. \label{fig:cofeb-system}} 
\end{figure}

Figure \ref{fig:cofeb-system} shows the geometry of the system under consideration. It is composed of a film of thickness 30\,nm made of a CoFeB alloy~\cite{Conca2014} with static magnetisation $M\tsub S = 1270$~kA/m and exchange constant $A \equiv \mu_0 M\tsub S^2 l^2/2 = 15$~pJ/m and a stripe of the same thickness made of permalloy with $M\tsub S = 760$~kA/m and exchange constant $A = 13$~pJ/m, separated from the film by a non-magnetic gap of thickness 10~nm. The stripe width~$w$ will be varied in the calculation described below. The gyromagnetic coefficient of both materials is taken to be $\gamma = -176$~GHz/T and the damping coefficient $\alpha = 0.0002$. The whole system is placed in a uniform external magnetic field $\mu_0 H = 0.1$~T directed along the negative $y$~axis and parallel to the stripe. 

\subsection{Eigenmodes}
\label{sec:eigenmodes}

\begin{figure}
 \centerline{\includegraphics{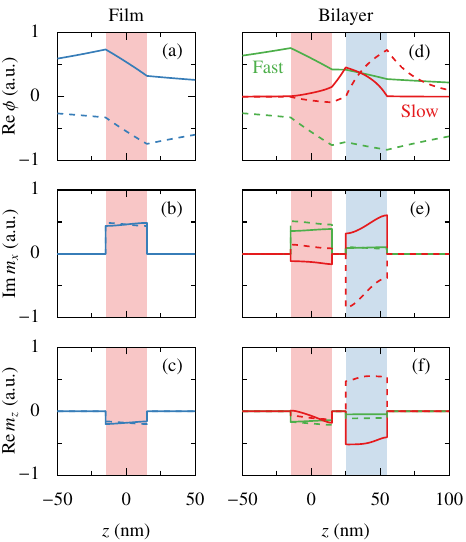}}
 \caption{Profiles of the magnetostatic potential $\phi(z)$ and the $x$ and $z$~components of the dynamic magnetisation $\vect m(z)$ of the eigenmodes of (a)--(c) the CoFeB film and (d)--(f) the bilayer. Solid lines: right-propagating modes; dashed lines: left-propagating modes. The areas taken by the CoFeB film and the permalloy film are shaded in red and blue, respectively. All plots show only the dominant real or imaginary component; the $L^2$ norm of the other one is over 100~times smaller.\label{fig:mode-profiles}} 
\end{figure}

Evidently, this system is composed of three $x$-invariant segments, two of which (the first and the third) are identical. The eigenmodes of each segment are calculated in the manner described in Sec.~\ref{sec:determination-of-eigenmodes}: Eqs.~\eqref{eq:weak-form} are discretised and turned into an algebraic generalised eigenvalue problem [Eq.~\eqref{eq:gep}] by expanding the fields $\phi$, $\tilde m_x$ and~$m_z$ into fifth-order Lagrange finite elements defined on a 1D mesh covering an interval of length $12.19$~\textmu m with the film at the centre. The same mesh is used in all three $x$-invariant segments. Mesh nodes are distributed so that the mesh is geometry-conforming in all segments; node spacing increases away from the ferromagnetic films. Dirichlet boundary conditions are imposed on $\phi$ at the top and bottom of the computational domain. In total, 224 degrees of freedom are used for $\phi$ and 6 degrees of freedom per ferromagnetic layer for $\tilde m_x$ and~$m_z$.

All calculations are done at the frequency $17$~GHz. At this frequency, we find that the CoFeB film supports a pair of counter-propagating propagative eigenmodes with wavelength 1020~nm and (amplitude) attenuation length 123~\textmu m. The CoFeB-permalloy bilayer supports two pairs of counter-propagating propagative eigenmodes; those propagating to the right have wavelengths 1299 and 108~nm and attenuation lengths 165 and 39~\textmu m, whereas those propagating to the left have wavelengths 973 and 145~nm and attenuation lengths 99 and 27~\textmu m. These values agree (to the number of digits shown) with ones obtained with the method of Wolfram and de Wames \cite{DeWames1970}, which does not require any domain truncation or discretisation.

The bilayer eigenmodes with wavelengths 1299 and 973~nm are concentrated primarily in the CoFeB layer and have larger group velocities than the modes with wavelengths 108 and 145 nm, concentrated in the permalloy layer. Therefore in the following, we shall call the former pair of modes the \emph{fast modes} and the latter the \emph{slow modes}. 

The $\phi(z)$, $m_x(z)$ and~$m_z(z)$ profiles of all propagative modes of the film and the bilayer are plotted in Fig.~\ref{fig:mode-profiles}. Here and throughout the rest of this paper, the phases of all mode profiles are chosen so that $m_z$ is real and negative at the midplane of the CoFeB film.

\begin{figure*}
 \centerline{\includegraphics{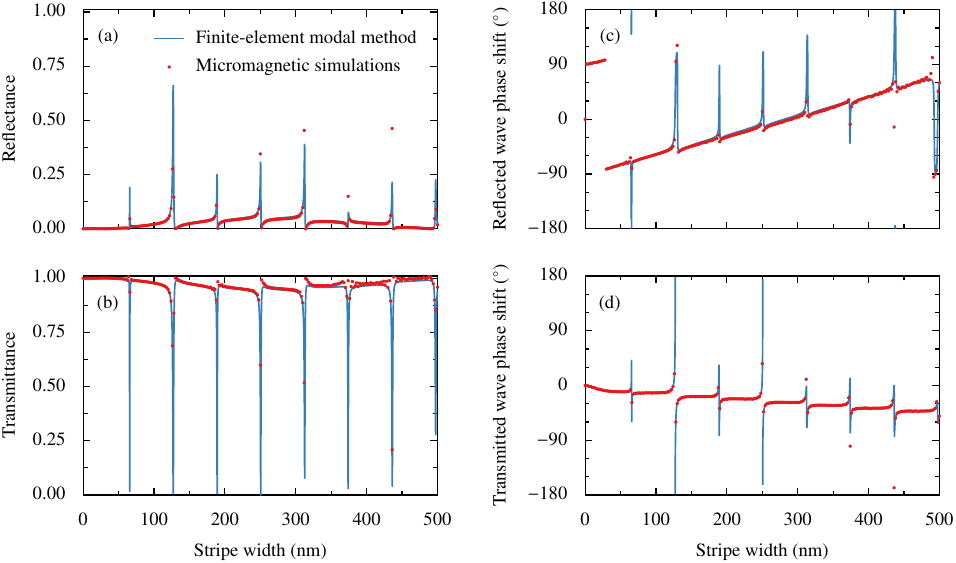}}
 \caption{Dependence of the scattering coefficients of the bilayer, obtained from numerical simulations done with the finite-element modal method and micromagnetic simulations, on the stripe width. All calculations were done at frequency 17~GHz.\label{fig:scattering-coeffs}} 
\end{figure*}

\subsection{Scattering: numerical simulations}

Suppose a right-propagating mode of the CoFeB film is excited by an antenna located to the left of segment~2. In that case, it will be scattered on the bilayer, giving rise to a reflected mode propagating to the left along segment~1 and a transmitted mode propagating to the right along segment~3. We are interested in the dependence of the power and phase of the reflected and transmitted modes on the width of the bilayer. We calculate the scattering coefficients in the manner described in Sec.~\ref{sec:solution-of-scattering-problem}, setting $\vect u^1$ to $[1, 0, 0, \dots]\transposed$ (i.e.\ assuming the incident field in segment 1 consists solely of its right-propagating propagative eigenmode with unit power, arriving at the interface between segments 1 and 2 with phase 0$^\circ$) and $\vect d^3$ to $[0, 0, \dots]\transposed$ (i.e.\ assuming there is no wave incident from the right in segment 3). The results of these calculations are plotted in Fig.~\ref{fig:scattering-coeffs} (solid lines). The four subplots show the reflectance and transmittance ($\abs{d_1^1/u_1^1}^2$ and $\abs{u_1^3/u_1^1}^2$, respectively) and the phase shifts of the reflected and transmitted waves, defined as $\arg(d_1^1/u_1^1)$ and $\arg\{u_1^3/[u_1^1 \exp(\I k_{x1}^{1u} w)]\}$. (The phase shift of the transmitted wave is defined as the difference of the phase of the transmitted wave and the phase that would be acquired by the incident wave if the stripe was removed.)  As mentioned at the end of Sec.~\ref{sec:power-flow}, in the presence of damping, waveguide eigenmodes are not strictly power-orthogonal; however, in the system under consideration the damping is small and the total cross-power (the sum of the terms proportional to $P_{ij}$, $i \neq j$, in the expansion from Eq.~\eqref{eq:power-expansion}) on both sides of the stripe never rises above 1\% of the incident power, so we neglect it in the following discussion.

On the reflectance curve from Fig.\ \ref{fig:scattering-coeffs}(a), we can see a regularly spaced series of narrow asymmetric peaks followed by zero crossings (the familiar Fano resonance shape), superimposed on a slow oscillation with a period of ca.~500\,nm. As expected from the energy conservation principle, the transmittance curve in Fig.\ \ref{fig:scattering-coeffs}(b) is a mirror image of the reflectance curve. The narrow peaks and dips in the reflectance and transmittance curves are accompanied by rapid changes of the phase shifts of the reflected and transmitted waves. Away from these narrow features, the phase shift of the transmitted wave decreases steadily with increasing stripe width, indicating that the phase of the transmitted wave lags more and more behind that of the unscattered incident wave.

\begin{figure}
 \centerline{\includegraphics{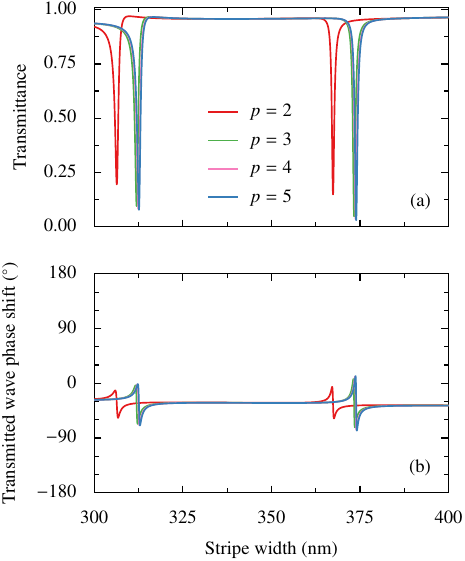}}
 \caption{Convergence of (a) the transmittance and (b) the phase shift of the transmitted wave with increasing polynomial degree~$p$ of the finite elements used in calculations.\label{fig:convergence}}
\end{figure}

The red circles in Fig.~\ref{fig:scattering-coeffs} are data points obtained from micromagnetic simulations performed with mumax$^3$ (Ref.~\onlinecite{vansteenkiste2014design}), described in detail in Appendix \ref{sec:micromagnetic-simulations}. The results of these simulations agree well with those obtained with the finite-element modal method. The latter can produce a highly precise solution to the linearised Landau-Lifshitz equations \eqref{eq:linearised-equations}. Figure~\ref{fig:convergence} shows the effect of increasing the polynomial degree of elements on the positions and shapes of two narrow features in the curves from Fig.\ \ref{fig:scattering-coeffs}(b) and (d). The curves obtained with elements of order 4 and~5 are visually almost indistinguishable. We have also verified that increasing the size of the computational domain in the $z$~direction by a factor of~4 makes no perceptible difference to the shape of the curves. 
Computation of the scattering coefficients of the system under consideration (using the mesh and element order described in Sec.~\ref{sec:eigenmodes}) with the finite-element modal method on a laptop PC takes 1.5~s. Over 97\% of this time is spent on the calculation of waveguide eigenmodes and interface scattering matrices, which needs to be done only once even if the scattering coefficients are to be computed for multiple stripe widths; the method is therefore particularly well suited for the modelling of structures containing waveguide segments whose lengths are allowed to vary. In contrast, micromagnetic simulations of the same system take approximately one hour for each value of~$w$. 

\subsection{Scattering: semi-analytical model}
\label{sec:model}

To understand the origin of the features visible in the plots from Fig.~\ref{fig:scattering-coeffs}, we formulate a semi-analytical model similar to that presented in Ref.~\onlinecite{Sobucki2021} for a system with segment 3 containing no magnetic materials. We start by noting that wave scattering on an interface $x = x^{i+1}$ separating segments $i$ and $i + 1$ can be described by a scattering matrix~$\mat S^{i+1}$ linking the complex amplitudes of the incoming and outgoing modes on both sides of the interface,
\begin{equation}
 \begin{bmatrix}
  \vect D^i(x^{i+1}) \\ \vect U^{i+1}(x^{i+1})
 \end{bmatrix}
 =
 \mat S^{i+1}
 \begin{bmatrix}
  \vect U^i(x^{i+1}) \\ \vect D^{i+1}(x^{i+1})
 \end{bmatrix}
 .
\end{equation}
This matrix can be easily calculated using the finite-element modal method; in the notation of Eq.~\eqref{eq:mode-match}, 
\begin{equation}
 \mat S^{i+1} = 
 \begin{bmatrix}
  -\mat V^{iu} & \mat W^{i+1,d}
 \end{bmatrix}
 ^{-1}
 \begin{bmatrix}
  \mat V^{id} & -\mat W^{i+1,u}
 \end{bmatrix}
 . 
\end{equation}

If segments $i$ and $i+1$ are long enough, all incoming evanescent modes decay away and become negligible before reaching the interface between these segments. To obtain the amplitudes of the outgoing propagative modes it is therefore sufficient to consider only the rows and columns of $\mat S^{i+1}$ corresponding to propagative modes. 

Consider first the interface $x = x^2$ at the left end of the bilayer. To simplify the notation, let us denote with $u\tsub i$ and $d\tsub i$ the complex amplitudes of the right- and left- propagating modes of the input film (segment~1) and with $u\tsub s$ and $d\tsub s$ ($u\tsub f$ and $d\tsub f$) the amplitudes of the right- and left-propagating slow (fast) modes of the bilayer (segment~2), all measured at $x = x^2$. If the bilayer is wide enough for the evanescent coupling between its ends to be negligible, then
\begin{equation}
 \label{eq:interface}
 \begin{bmatrix}
  d\tsub i \\ u\tsub s \\ u\tsub f
 \end{bmatrix}
 =
 \begin{bmatrix}
  S\tsub{ii} & S\tsub{is} & S\tsub{if} \\
  S\tsub{si} & S\tsub{ss} & S\tsub{sf} \\
  S\tsub{fi} & S\tsub{fs} & S\tsub{ff}
 \end{bmatrix}
 \begin{bmatrix}
  u\tsub i \\ d\tsub s \\ d\tsub f
 \end{bmatrix}
 ,
\end{equation}
where $S\tsub{ii}$ etc.\ are appropriate elements of the scattering matrix $\mat S^2$. At 17~GHz, their numerical values found with the finite-element modal method are 
\begin{multline}
 \begin{bmatrix}
  S\tsub{ii} & S\tsub{is} & S\tsub{if} \\
  S\tsub{si} & S\tsub{ss} & S\tsub{sf} \\
  S\tsub{fi} & S\tsub{fs} & S\tsub{ff}
 \end{bmatrix}
 \\
 = 
 \begin{bmatrix}
0.117 \E^{-0.04\I} & 
0.089 \E^{-0.78\I} & 
0.989 \E^{0.03\I} 
\\ 
0.145 \E^{-1.35\I} & 
0.984 \E^{2.95\I} & 
0.095 \E^{0.80\I} 
\\ 
0.983 \E^{-0.05\I} & 
0.149 \E^{1.17\I} & 
0.111 \E^{-3.00\I} 
 \end{bmatrix}
\end{multline}
(these values are obtained for modes normalised to carry unit power, with phases chosen so that $m_z$ is real and negative on the mid-plane of the CoFeB film). It can be seen that the film mode is coupled primarily with the fast mode of the bilayer. The slow bilayer mode is strongly reflected. The fast and slow bilayer modes are only weakly coupled. 

Likewise, amplitudes of the incoming and outgoing modes at the right end of the bilayer ($x = x^3$) are tied by
\begin{equation}
 \label{eq:interface'}
 \begin{bmatrix}
  d'\tsub s \\ d'\tsub f \\ u'\tsub o
 \end{bmatrix}
 =
 \begin{bmatrix}
  S'\tsub{ss} & S'\tsub{sf}
  \\
  S'\tsub{fs} & S'\tsub{ff}
  \\
  S'\tsub{os} & S'\tsub{of}
 \end{bmatrix}
 \begin{bmatrix}
  u'\tsub s \\ u'\tsub f 
 \end{bmatrix}
 ,
\end{equation}
where $u\tsub s'$ and $d\tsub s'$ ($u\tsub f'$ and $d\tsub f'$) are the amplitudes of the right- and left-propagating slow (fast) modes of the bilayer and 
$u\tsub o'$ is the amplitude of the right-propagating mode of the CoFeB film, 
all measured at $x = x^3$. Numerically \footnote{Owing to the geometrical symmetry of the system, 
$S'_{kl} \approx S_{lk}$ and $S'_{\textrm{o}k} \approx S_{k\textrm{i}}$ for $k, l = \textrm{f}, \textrm{s}$. The equality would be exact if the modes were orthonormalised with respect to the unconjugated inner product defined in Eq.~\eqref{eq:lorentz-ortho-same-unconjugated} rather than normalised to unit power.},
\begin{equation}
  \begin{bmatrix}
  S'\tsub{ss} & S'\tsub{sf}
  \\
  S'\tsub{fs} & S'\tsub{ff}
  \\
  S'\tsub{os} & S'\tsub{of}
  \end{bmatrix} =
\begin{bmatrix}
0.984 \E^{2.95\I} & 
0.149 \E^{1.17\I} 
\\ 
0.095 \E^{0.80\I} & 
0.111 \E^{-3.00\I} 
\\ 
0.145 \E^{-1.34\I} & 
0.983 \E^{-0.05\I} 
\end{bmatrix}.
\end{equation}
Mode amplitudes at the two ends of the bilayer are linked by 
\begin{subequations}
 \label{eq:resonator}
 \begin{align}
  u_i' &= \exp(\I k_{iu} w)\, u_i \eqqcolon \Phi_{iu} u_i,\\
  d_i &= \exp(-\I k_{id} w)\, d_i' \eqqcolon \Phi_{id} d_i'\quad\text{for}\quad i = \text{s}, \text{f},
 \end{align}
\end{subequations}
where $k_{iu}$ and $k_{id}$ are the wave numbers of the right- and left-propagating modes, numerically determined to be $k_{\text{s}u} = 58.3 + 0.026\I$, $k_{\text{f}u} = 4.84 + 0.006\I$, $k_{\text{s}d} = -43.4 - 0.037\I$ and $k_{\text{f}d} = -6.46 - 0.010\I$~rad/\textmu m.

Together, Eqs.\ \eqref{eq:interface}, \eqref{eq:interface'} and \eqref{eq:resonator} 
form a system of ten equations for as many unknown mode amplitudes (the amplitude $u\tsub i$ of the mode incident from the input film is treated as known). To obtain intelligible expressions for the scattering coefficients $r \equiv d\tsub i / u\tsub i$ and  $t \equiv u\tsub o / u\tsub i$, it is advantageous to start by eliminating the amplitudes $u\tsub f$, $d\tsub f$, $u\tsub f'$ and $d\tsub f'$ of the fast bilayer mode, which is only weakly reflected at the interface with the CoFeB film and hence will not give rise to strong Fabry-Perot-like resonances. This mimics the approach taken by Lecamp \textit{et al.}~\cite{Lecamp} in their model of pillar microcavities. This reduces the second row of Eq.~\eqref{eq:interface} and the first row of Eq.~\eqref{eq:interface'} to
\begin{subequations}
\begin{align}
  u\tsub s &= \tilde S\tsub{si} u\tsub i + \tilde S\tsub{ss} d\tsub s, \\
  d\tsub s' &= \tilde S\tsub{ss}' u\tsub s' + \tilde S\tsub{sf}' \Phi_{\text{f}u} S\tsub{fi} u\tsub i,
\end{align}
\end{subequations}
where
\begin{subequations}
 \label{eq:S-tilde}
 \begin{align}
  \tilde S\tsub{si} 
  &\coloneqq 
  \frac{S\tsub{si} + \alpha\tsub f S\tsub{sf} \Phi_{\text{f}d} S\tsub{ff}' \Phi_{\text{f}u} S\tsub{fi}}
  {1 - \alpha\tsub f S\tsub{sf} \Phi_{\text{f}d} S\tsub{fs}' \Phi_{\text{s}u}},
  \\
  \tilde S\tsub{ss}
  &\coloneqq
  \frac{S\tsub{ss} + \alpha\tsub f S\tsub{sf} \Phi_{\text{f}d} S\tsub{ff}' \Phi_{\text{f}u} S\tsub{fs}}
  {1 - \alpha\tsub f S\tsub{sf} \Phi_{\text{f}d} S\tsub{fs}' \Phi_{\text{s}u}},
  \\
  \tilde S\tsub{ss}'
  &\coloneqq
  \frac{S\tsub{ss}' + \alpha\tsub f S\tsub{sf}' \Phi_{\text{f}u} S\tsub{ff} \Phi_{\text{f}d} S\tsub{fs}'}
  {1 - \alpha\tsub f S\tsub{sf}' \Phi_{\text{f}u} S\tsub{fs} \Phi_{\text{s}d}},
  \\
  \tilde S\tsub{sf}' 
  &\coloneqq
  \frac{\alpha\tsub f S\tsub{sf}'}
  {1 - \alpha\tsub f S\tsub{sf}' \Phi_{\text{f}u} S\tsub{fs} \Phi_{\text{s}d}}
 \end{align}
\end{subequations}
and 
\begin{equation}
 \alpha\tsub f \coloneqq (1 - S\tsub{ff} \Phi_{\text{f}d} S\tsub{ff}' \Phi_{\text{f}u})^{-1}.
\end{equation}
The fast bilayer mode is only weakly reflected at the interface with the film: $\abs{S\tsub{ff}} = \abs{S'\tsub{ff}} \approx 0.111 \ll 1$. Therefore multiple reflections of the fast mode at bilayer interfaces do not give rise to strong Fabry-Perot resonances and the coefficient $\alpha\tsub f$ remains close to 1 for all bilayer widths. Given that in addition all reflection coefficients except $S\tsub{ss}$ and $S\tsub{ss}'$ are small, we can expect the scattering coefficients with a tilde defined in Eq.~\eqref{eq:S-tilde} to be close to the corresponding coefficients without a tilde.

\begin{figure*}
 \centerline{\includegraphics{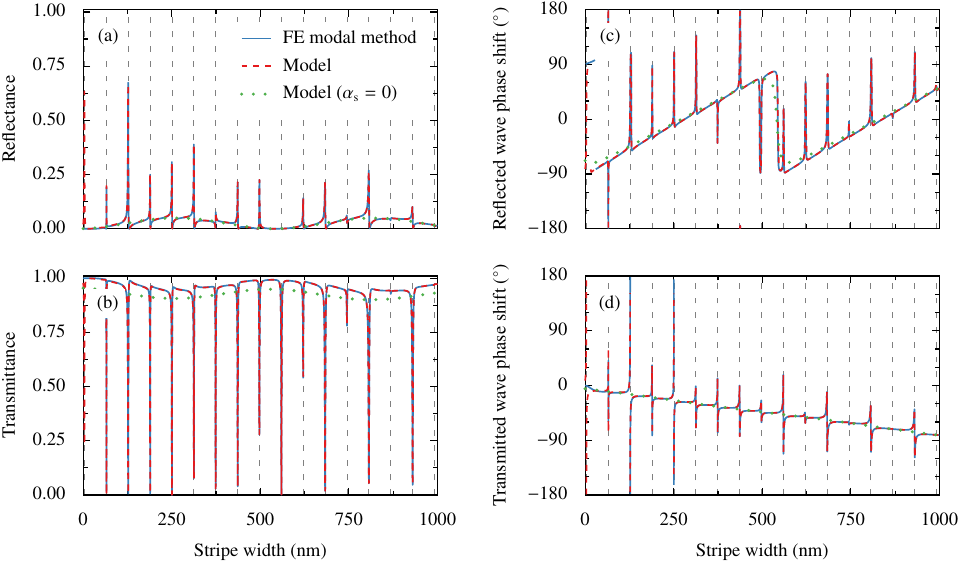}}
 \caption{Comparison of the values of the bilayer's scattering coefficients predicted by the model from Sec.~\ref{sec:model} with results of numerical simulations done with the finite-element modal method. The dashed vertical lines are the positions of resonances predicted with Eq.\ \eqref{eq:resonant-widths}. \label{fig:approximate-scattering-coeffs}}
\end{figure*}

Having eliminated the amplitudes of the fast modes, we solve the remaining equations for the amplitudes of the slow modes and substitute the resulting expressions into the formulas for $d\tsub i$ in the first row of Eq.~\eqref{eq:interface} and $u\tsub o'$ in the last row of Eq.~\eqref{eq:interface'}. This yields the following formulas for the reflection and transmission coefficients:
\begin{subequations}
 \label{eq:approx-r,t}
 \begin{align}
 \label{eq:approx-r}
 r &\equiv d\tsub i/u\tsub i = r\tsub f + \alpha\tsub s r\tsub s,\\
 \label{eq:approx-t}
 t &\equiv u\tsub o/u\tsub i = t\tsub f + \alpha\tsub s t\tsub s,
 \end{align}
\end{subequations}
where 
\begin{equation}
 \label{eq:beta}
 \alpha\tsub s \coloneqq (1 - \tilde S\tsub{ss} \Phi_{\text{s}d}
 \tilde S\tsub{ss}' \Phi_{\text{s}u})^{-1}
\end{equation}
represents the effect of multiple reflections of the slow mode and
\begin{subequations}
 \label{eq:r,t-terms}
 \begin{align}
  \label{eq:rf}
  r\tsub f &\coloneqq \smallfactor{S\tsub{ii}} + \alpha\tsub f S\tsub{if} \Phi_{\text{f}d} \smallfactor{S\tsub{ff}'} \Phi_{\text{f}u} S\tsub{fi},\\
  \begin{split}
  \label{eq:rs}
  r\tsub s &\coloneqq \smallfactor{S\tsub{is}} \Phi_{\text{s}d}(\tilde S\tsub{ss}' \Phi_{\text{s}u} \smallfactor{\tilde S\tsub{si}} +
  \smallfactor{\tilde S\tsub{sf}'} \Phi_{\text{f}u} S\tsub{fi}) \\
  &\hphantom{:}\quad + \alpha\tsub f S\tsub{if} \Phi_{\text{f}d} \bigl[
    \smallfactor{S\tsub{fs}'} \Phi_{\text{s}u} 
    (\smallfactor{\tilde S\tsub{si}} + \tilde S\tsub{ss} \Phi_{\text{s}d} \smallfactor{\tilde S\tsub{sf}'} \Phi_{\text{f}u} S\tsub{fi}) \\
    &\hphantom{:}\quad + \smallfactor{S\tsub{ff}'} \Phi_{\text{f}u} \smallfactor{S\tsub{fs}} \Phi_{\text{s}d} 
    (\tilde S\tsub{ss}' \Phi_{\text{s}u} \smallfactor{\tilde S\tsub{si}} + \smallfactor{\tilde S\tsub{sf}'} \Phi_{\text{f}u} S\tsub{fi}) 
    \bigr],
  \end{split}
  \\
  \label{eq:tf}
  t\tsub f &\coloneqq \alpha\tsub f S\tsub{of}' \Phi_{\text{f}u} S\tsub{fi},\\
  \begin{split}
  \label{eq:ts}
  t\tsub s &\coloneqq \smallfactor{S\tsub{os}'} \Phi_{\text{s}u}
  (\smallfactor{\tilde S\tsub{si}} +
  \tilde S\tsub{ss} \Phi_{\text{s}d} \smallfactor{\tilde S\tsub{sf}'}
  \Phi_{\text{f}u} S\tsub{fi}) \\
  &\hphantom{:}\quad + \alpha\tsub f S\tsub{of}' \Phi_{\text{f}u} \bigl[
    \smallfactor{S\tsub{fs}} \Phi_{\text{s}d} 
    (\smallfactor{\tilde S\tsub{sf}'} \Phi_{\text{f}u} S\tsub{fi} + 
     \tilde S\tsub{ss}' \Phi_{\text{s}u} \smallfactor{\tilde S\tsub{si}}) \\
    &\hphantom{:}\quad + \smallfactor{S\tsub{ff}} \Phi_{\text{f}d} \smallfactor{S\tsub{fs}'} \Phi_{\text{s}u} 
    (\smallfactor{\tilde S\tsub{si}} +
    \tilde S\tsub{ss} \Phi_{\text{s}d} \smallfactor{\tilde S\tsub{sf}'} \Phi_{\text{f}u} S\tsub{fi}) 
    \bigr]
  \end{split}  
 \end{align}
\end{subequations}
(to facilitate interpretation, scattering coefficients of magnitude much less than 1 have been underlined). It can be seen that both scattering coefficients are made up of two terms. 

The first term, $r\tsub f$ or $t\tsub f$, is free from the resonant factor  $\alpha\tsub s$ and the rapidly varying phase factors $\Phi_{\text{d}s}$ and $\Phi_{\text{u}s}$. Both terms in $r\tsub f$ contain one small reflection coefficient, whereas $t\tsub f$ contains none. Therefore the transmittance $\abs{t}^2$ is usually larger than the reflectance $\abs{r}^2$. 

The other term, $\alpha\tsub s r\tsub s$ or $\alpha\tsub s t\tsub s$, is proportional to the factor $\alpha\tsub s$, which is normally close to unity but its magnitude can grow to over~60 near Fabry-Perot resonances of the slow mode of the bilayer. These occur approximately at stripe widths
\begin{equation}
\label{eq:resonant-widths}
w_n = \frac{2\pi n - \arg (S\tsub{ss} S\tsub{ss}')}{\RE(k_{\text{s}u} + k_{\text{s}d})}
\quad
\text{where}
\quad
n = 1, 2, \dots
\end{equation}
(neglecting the small difference between $S\tsub{ss} S\tsub{ss}'$ and $\tilde S\tsub{ss} \tilde S\tsub{ss}'$). Away from these resonances, this term is small, since each of the terms making up $r\tsub s$ or $t\tsub s$ is proportional to a product of at least two scattering coefficients of small magnitude (less than 0.15). All these terms also contain the phase factors $\Phi_{\text{s}u}$ and/or $\Phi_{\text{s}d}$, so their phases vary rapidly.

In Fig.~\ref{fig:approximate-scattering-coeffs} we compare the power and phase shift of the reflected and transmitted modes calculated from Eq.~\eqref{eq:approx-r,t} (red solid curve) with results of full numerical simulations made with the finite-element modal method (black symbols). The two data series agree very well except for very narrow stripes, for which evanescent coupling between the two ends of the bilayer, neglected in the model from which Eq.~\eqref{eq:approx-r,t} was derived, plays a large role. The vertical lines indicate the positions of resonances predicted from Eq.~\eqref{eq:resonant-widths}; their agreement with the full numerical simulations justifies approximating $\tilde S\tsub{ss} \tilde S\tsub{ss}'$ with $S\tsub{ss} S\tsub{ss}'$ in the derivation of that equation.

The green dotted curves in Fig.~\ref{fig:approximate-scattering-coeffs} show the result of neglecting the terms proportional to $\alpha\tsub s$ in Eq.~\eqref{eq:approx-r,t}. As expected, the sharp resonances are gone; however, the curves continue to reproduce faithfully long-term trends. Thus, the slow oscillations of the reflectance and transmittance as a function of stripe width are due to the weak Fabry-Perot resonances of the fast mode encapsulated in the $\alpha\tsub f$ factor. The propagation constant of the right-propagating fast mode is smaller than that of the eigenmodes of the CoFeB film, hence the phase shift of the transmitted wave decreases steadily with stripe width.

\section{Conclusions}

We have introduced a finite-element modal method for the simulation of spin waves in the dipole-exchange regime and the Damon-Eshbach configuration. We have complemented it with a derivation of the Lorentz reciprocity theorem and mode orthogonality relations applicable to this class of systems and extended the formula for the power carried by magnetostatic modes to the case of dipole-exchange spin waves. We have used a system composed of a CoFeB thin film decorated with a dynamically coupled Py stripe to illustrate the usefulness of the proposed method for the calculation of spin-wave transmittance, reflectance, and the phase shift of scattered waves. Its predictions were successfully validated against micromagnetic simulations. We found the calculation of scattering coefficients with the new method to be over 1000 times faster than with micromagnetic simulations, clearly demonstrating its potential in the design of elements of magnonic circuits. Table \ref{tab:comparison} highlights the main differences between the modal method and micromagnetic simulations.

\begin{table*}
\begin{ruledtabular}
\footnotetext[1]{Segments with variable cross-sections or non-negligible spatial variability of static magnetisation could, however, be discretised with standard non-modal finite elements, as in photonics \cite[Sec.\ 11.1.3]{Jin2015}.}
\begin{tabular}{>{\raggedright}p{0.2\textwidth}p{0.35\textwidth}p{0.35\textwidth}}
\textbf{Aspect} &
\textbf{Modal method} &
\textbf{Micromagnetic simulations} \\
Physically relevant outputs (scattering coefficients and matrices) &
Obtained directly without any post\-process\-ing. &
Require postprocessing (see Appendix \ref{sec:micromagnetic-simulations}). \\
Outgoing boundary conditions in input and output waveguides &
Fulfilled automatically. &
Fulfilment requires introducing boundary layers with a suitable damping profile.\\
Individual mode excitation &
Straightforward (via the $\vect u^1$ and $\vect d^n$ amplitude vectors). &
Non-trivial (requires designing an appropriate antenna) unless the mode to be excited is the only one with a specific symmetry.\\
Number of unknowns &
Low (a few hundreds per segment). &
High (864 million for the three-segment system simulated in this paper). \\
Simulation time &
Low (1 s per point for the system studied in this paper). &
High (1 hour per point for the system studied in this paper). \\
Potential for reuse of pre-calculated results &
Segment eigenmodes can be pre-calculated and reused in simulations of multiple systems containing segments with the same cross-section.
Simulations at each frequency must be done separately. &
Each change of geometry requires running the simulation from scratch. 
A single simulation with a wide-band source provides information about system behaviour at multiple frequencies. \\
Precision &
High: only a modest computational cost is required to reduce numerical errors (caused by the finite mesh size and density and finite polynomial expansion order) far below those resulting from the adopted mathematical model (e.g.\ linearisation of the LL equation, piecewise constant material properties, idealised geometry). &
Typically low: numerical errors are difficult to eliminate especially when lowest-order finite differences are employed. \\
Temporal behaviour &
Time-harmonic field variation assumed and exploited to simplify and accelerate computations. &
Not limited to time-harmonic field variation but unable to exploit it if present (requires time integration until steady state). \\
Nonlinearity of the LL equation &
Neglected (small perturbations regime assumed). &
Fully taken into account. \\
System geometry  &
Limited to systems that can be split into segments with constant cross-sections.\footnotemark[1] &
Arbitrary. \\
Static magnetisation &
Required to be spatially piecewise constant in terms of both magnitude and orientation.\footnotemark[1] &
Arbitrary. \\
\end{tabular}
\end{ruledtabular}
 \caption{Comparison between the finite-element modal method and micromagnetic simulations.\label{tab:comparison}}
\end{table*}

We have formulated a detailed semi-analytical model of spin-wave propagation in the system mentioned above. The only numerical inputs required by the model, namely propagation constants of individual waveguide modes and normalised mode scattering coefficients associated with interfaces separating waveguide segments with different geometry, were obtained directly with the modal method (with no post-processing required). The model highlights the contrasting roles played by the two pairs of normal modes supported by the bilayered part of the system and makes it possible to quantify the contributions of individual scattering pathways. In particular, we have found the slow modes to be responsible for the formation of sharp Fabry-Perot-like resonances observed in the transmission spectra, as in the Gires–Tournois interferometer \cite{Sobucki2021}, and the fast modes to account for low-amplitude transmittance oscillations with a larger spatial period. This provides a deepened physical interpretation of the results of recent experiments \cite{Qin2021} and a tool for future research and optimisation of magnonic devices.

\begin{acknowledgments}
  The research leading to these results has received funding from the National Science Centre of Poland, project no. 2019/35/D/ST3/03729 (PG and KS), and 2018/30/Q/ST3/00416 (MK). 
  The simulations were partially performed at the Poznan Supercomputing and Networking Center (Grant No. 398). \copyright\ Crown Copyright, Met Office 2023.
\end{acknowledgments}

\appendix

\section{Mode matching matrices}
\label{sec:mode-matching-matrices}

Let $\{f_q(z)\}_{q=1}^{N^f}$ be the set of (continuous) basis functions used to expand the magnetostatic potential profiles of the eigenmodes of all waveguide segments. Let $\{g_q^{\mathcal{L}}(z)\}_{q=1}^{N^g(\mathcal{L})}$ be the (possibly empty) set of basis functions obtained by restricting all basis functions $f_q(z)$ to a set of intervals $\mathcal{L} \subset [z\tsub{min}, z\tsub{max}]$ and keeping only those that are not identically zero. Let $\mathcal{L}^i \subset [z\tsub{min}, z\tsub{max}]$ be the set of intervals with nonvanishing $l^2(z)$ in the $i$th segment. The set $\{g_q^{\mathcal{L}^i}(z)\}_{q=1}^{N^g(\mathcal{L}^{i})}$ is then the set of basis functions used to expand the magnetisation profiles of the eigenmodes of the $i$th waveguide segment.

The solution of the eigenproblem~\eqref{eq:gep} for each segment~$i$ yields a family of $2[N^f + 2 N^g(\mathcal{L}^i)]$ eigenmodes; the field profiles of the $j$th mode propagating or decaying to the right with wavenumber $k_{xj}^{iu}$ are
\begin{subequations}
 \label{eq:mode-field-profiles}
 \begin{align}
  \phi_j^{iu}(z) &= \sum_{q=1}^{N^f} f_q(z) F_{qj}^{iu}, \\
  \tilde m_{xj}^{iu}(z) &= \sum_{q=1}^{N^g(\mathcal{L}^i)} 
   g_q^{\mathcal{L}^i}(z) \tilde M_{xqj}^{iu}, \\
  m_{zj}^{iu}(z) &= \sum_{q=1}^{N^g(\mathcal{L}^i)} 
   g_q^{\mathcal{L}^i}(z) M_{zqj}^{iu},
 \end{align}
\end{subequations}
where $F_{qj}^{iu}$, $\tilde M_{xqj}^{iu}$ and $M_{zqj}^{iu}$ are elements of one of the eigenvectors (optionally scaled to normalise the mode to unit power). Replacement of the superscript~$u$ with~$d$ yields analogous expressions for the $j$th mode propagating or decaying to the left.

Imposition of the boundary conditions listed in Sec.~\ref{sec:mode-matching} at the interface $x=x^{i+1}$ between segments $i$ and $i+1$ produces Eq.~\eqref{eq:mode-match} with $\mat E^{iu}(\Delta x)$ defined as the diagonal matrix whose $j$th diagonal element is $\exp(\I k_{xj}^{iu} \Delta x)$ and the $\mat V^{iu}$ and $\mat W^{i+1,u}$ matrices defined as
\begingroup
\renewcommand*{\arraystretch}{1.25}
\begin{equation}
  \mat V^{iu} = 
  \begin{bmatrix}
   \mat J^{\phi\phi} \mat F^{iu} \\
    \mat J^{\phi m_i} \tilde{\mat M}_x^{iu} - 
     \I \mat J^{\phi\phi} \mat F^{iu} \mat K_x^{iu} \\
    \mat J^{m_{i \cap i+1} m_i} \tilde{\mat M}_x^{iu} \\
    \mat J^{m_{i \cap i+1} m_i} \mat M_z^{iu} \\
    \mat J_{l^2}^{m_{i \cup i+1} m_i} \tilde{\mat M}_x^{iu} \\
    \mat J_{l^2}^{m_{i \cup i+1} m_i} \mat M_z^{iu} 
  \end{bmatrix}
\end{equation}
and
\begin{equation}
 \mat W^{i+1,u} = 
 \begin{bmatrix}
  \mat J^{\phi\phi} \mat F^{i+1,u} \\
   \mat J^{\phi m_{i+1}} \tilde{\mat M}_x^{i+1,u} - 
    \I \mat J^{\phi\phi} \mat F^{i+1,u} \mat K_x^{i+1,u} \\
   \mat J^{m_{i \cap i+1} m_{i+1}} \tilde{\mat M}_x^{i+1,u} \\
   \mat J^{m_{i \cap i+1} m_{i+1}} \mat M_z^{i+1,u} \\
   \mat J_{l^2}^{m_{i \cup i+1} m_{i+1}} \tilde{\mat M}_x^{i+1,u} \\
   \mat J_{l^2}^{m_{i \cup i+1} m_{i+1}} \mat M_z^{i+1,u} 
 \end{bmatrix}
 .
\end{equation}
\endgroup
In the equations above, $\mat F^{iu}$, $\tilde{\mat M}_x^{iu}$ and $\mat M_z^{iu}$ are matrices of the mode field expansion coefficients $F_{qj}^{iu}$, $\tilde M_{xqj}^{iu}$ and $M_{zqj}^{iu}$ introduced in Eq.~\eqref{eq:mode-field-profiles}, whereas the elements of matrices $\mat J^{\cdots}$ are defined as
\begin{subequations}
 \begin{align}
  J_{pq}^{\phi\phi} &= \int f_p(z)\,f_q(z)\,\diff z,\\
  J_{pq}^{\phi m_j} &= \int f_p(z)\,g_q^{\mathcal{L}^j}(z)\,\diff z,\\
  J_{pq}^{m_{i \cap i+1} m_j} &= \int g_p^{\mathcal{L}^i \cap \mathcal{L}^{i+1}}(z)\,g_q^{\mathcal{L}^j}(z)\,\diff z,\\
  J_{l^2,pq}^{m_{i \cup i+1} m_j} &= \int l_j^2(z) \, g_p^{\mathcal{L}^i \cup \mathcal{L}^{i+1}}(z)\,g_q^{\mathcal{L}^j}(z)\,\diff z,  
 \end{align}
\end{subequations}
with $l_j^2(z)$ representing the profile of the squared exchange length in segment $j$. Finally, $\mat K_x^{iu}$ is the diagonal matrix of mode wavenumbers $k_{xj}^{iu}$. The formulas for $\mat E^{id}(\Delta x)$, $\mat V^{id}$ and $\mat W^{i+1,d}$ can be obtained by replacing the superscript $u$ with $d$.

\section{Mode orthogonality relations}
\label{sec:ortogonality-relations}

In this Appendix, we derive a version of the Lorentz reciprocity theorem applicable to dipole-exchange spin waves in the Damon-Eshbach configuration and a number of orthogonality relations binding pairs of eigenmodes of such structures. In this paper, these relations are utilised to deduce the formula for mode power [Eq.~\eqref{eq:mode-power}]. However, they can also be useful in their own right, for instance to extract the contribution of a particular mode to the total magnetisation calculated with a non-modal method \cite{McIsaac1991}.

\subsection{Lorentz reciprocity theorem for dipole-exchange spin waves}

Consider a magnetostatic potential $\phi_a$ and magnetisation $\vect m_a$ satisfying the system of equations \eqref{eq:linearised-equations}, comprising the Gauss law for magnetism and the linearised LL equation with a damping term, which can be rewritten in the following form:
\begin{subequations}
 \label{eq:gauss-ll-a}
 \begin{align}
  \label{eq:gauss-a}
  \vect\nabla \cdot (\vect m_a - \vect\nabla \phi_a) &= 0,\\
  \begin{split}
  \label{eq:ll-a}
  \vect\nabla \phi_a 
  - \sum_{i=x,z} \vect e_i [\vect\nabla \cdot (l^2 \vect\nabla m_{ai})]
  \\
  + \frac{\omega_0 - \I\omega\alpha}{\omega_M} \vect m_a - \I \frac{\omega}{\omega_M}\, \vect e_y \times \vect m_a &= 0,
  \end{split}
 \end{align}
\end{subequations}
where $\vect e_i$ ($i = x, y, z$) denotes the unit vector directed along axis $i$. Consider also another magnetostatic potential $\phi_b'$ and magnetisation $\vect m_b'$ satisfying the corresponding equations in the complementary system, i.e.\ one obtained by reversing the direction of the static external magnetic field and the static magnetisation and replacing damping with gain:
\begin{subequations}
 \label{eq:gauss-ll-b}
 \begin{align}
  \label{eq:gauss-b}
  \vect\nabla \cdot (\vect m_b' - \vect\nabla \phi_b') &= 0,\\
  \begin{split}
  \label{eq:ll-b}
  \vect\nabla \phi_b'
  - \sum_{i=x,z} \vect e_i [\vect\nabla \cdot (l^2 \vect\nabla m_{bi}')]
  \\
  + \frac{\omega_0 - \I\omega\alpha}{\omega_M} \vect m_b' + \I \frac{\omega}{\omega_M}\, \vect e_y \times \vect m_b' &= 0.
  \end{split}
 \end{align}
\end{subequations}
Multiplying Eq.~\eqref{eq:gauss-a} by $\phi_b'$ and Eq.~\eqref{eq:gauss-b} by $\phi_a$ and subtracting the results, we obtain
\begin{equation}
  \label{eq:sub-gauss}
  \phi_b' \vect\nabla \cdot \vect m_a - \phi_a \vect\nabla \cdot \vect m_b'
  - \phi_b' \nabla^2 \phi_a + \phi_a \nabla^2 \phi_b' = 0.  
\end{equation}
Similarly, multiplying Eq.~\eqref{eq:ll-a} by $\vect m_b'$ and Eq.~\eqref{eq:ll-b} by $\vect m_a$ and subtracting the results, we obtain
\begin{multline}
  \label{eq:sub-ll}
  \vect m_b' \cdot \vect\nabla \phi_a 
  - \vect m_a \cdot \vect\nabla \phi_b'
  - \sum_{i=x,z} m_{bi}' \vect\nabla \cdot (l^2 \vect\nabla m_{ai})
  \\
  + \sum_{i=x,z} m_{ai} \vect\nabla \cdot (l^2 \vect\nabla m_{bi}') = 0.
\end{multline}
Subtraction of Eq.~\eqref{eq:sub-gauss} from Eq.~\eqref{eq:sub-ll} yields
\begin{multline}
  \vect m_b' \cdot \vect\nabla \phi_a 
  - \vect m_a \cdot \vect\nabla \phi_b'
  - \phi_b' \vect\nabla \cdot \vect m_a + \phi_a \vect\nabla \cdot \vect m_b'
  \\
  + \phi_b' \nabla^2 \phi_a - \phi_a \nabla^2 \phi_b' 
  - \sum_{i=x,z} m_{bi}' \vect\nabla \cdot (l^2 \vect\nabla m_{ai})
  \\
  + \sum_{i=x,z} m_{ai} \vect\nabla \cdot (l^2 \vect\nabla m_{bi}') = 0.
\end{multline}
Using the relationship $\vect b = \mu_0(\vect m + \vect h) = \mu_0(\vect m - \vect\nabla \phi)$ and the identity $\vect\nabla \cdot (f \vect g) = (\vect\nabla f) \cdot \vect g + f \vect\nabla \cdot \vect g$, this equation can be reduced to
\begin{multline}
 \label{eq:lorentz-theorem}
 \nabla \cdot [\mu_0^{-1} (\phi_a \vect b_b' - \phi_b' \vect b_a) 
 \\
  +\sum_{i=x,z} l^2 (m_{ai} \vect\nabla m_{bi}' - m_{bi}' \vect\nabla m_{ai})] = 0.
\end{multline}
This is an analogue of the Lorentz reciprocity theorem, known from classical electromagnetism \cite{Harrington}, for magnetostatic waves with exchange interaction. As in electromagnetism \cite{Villeneuve}, it can be used to derive orthogonality relations for waveguide modes.

\subsection{Orthogonality relation between modes of complementary waveguides}

Let $\phi_a$ and $\vect m_a$ be the magnetostatic potential and magnetisation of an eigenmode of an $x$-invariant waveguide, with $k_{xa}$ being the mode wavenumber, and $\phi_b'$ and $\vect m_b'$ the corresponding fields of an eigenmode of the complementary waveguide, with wavenumber $k_{xb}'$. Applying the Lorentz reciprocity theorem \eqref{eq:lorentz-theorem} to these fields and taking advantage of the fact that they can be written as a product of $\exp(\I k_{xa} x)$ or $\exp(\I k_{xb}' x)$ and a $z$-dependent factor, we get
\begin{multline}
 \I (k_{xa} + k_{xb}')[\mu_0^{-1} (\phi_a b_{bx}' - \phi_b' b_{ax}) -
  \I l^2 (k_{xa} - k_{xb}') \vect m_{a} \cdot \vect m_{b}'] 
  \\
 +\partial_z[ \mu_0^{-1} (\phi_a b_{bz}' - \phi_b' b_{az}) + 
  l^2 (\vect m_{a} \, \partial_z \vect m_{b}' - \vect m_{b}' \, \partial_z \vect m_{a})]
  = 0.
\end{multline}
Integrating over $z$ and noting that $\phi$, $b_z$ and $\vect m \cdot l^2 \partial_z \vect m$ are continuous functions of $z$ and (at least when any magnetic layers have finite thickness) $\phi$ and $\vect m$ decay to 0 as $z \to \pm\infty$, we see that the second term $\int_{-\infty}^\infty \partial_z(\ldots) \, \diff z$ vanishes, leaving us with
\begin{multline}
 (k_{xa} + k_{xb}') \int_{-\infty}^\infty [\mu_0^{-1} (\phi_a b_{bx}' - \phi_b' b_{ax}) 
 \\
 -\I l^2 (k_{xa} - k_{xb}') \vect m_{a} \cdot \vect m_{b}']\,\diff z = 0.
\end{multline}
This implies the following orthogonality relation:
\begin{multline}
 \label{eq:lorentz-ortho-complementary}
 \int_{-\infty}^\infty [\mu_0^{-1} (-\phi_a b_{bx}' + \phi_b' b_{ax}) 
 + \I l^2 (k_{xa} - k_{xb}') \vect m_{a} \cdot \vect m_{b}']\,\diff z = 0
 \\
  \text{if}\quad
  k_{xa} \neq -k_{xb}'.
\end{multline}

\subsection{Orthogonality relation between modes of a single waveguide (without conjugation)}

Let [$\phi_a(z)$, $m_{ax}(z)$, $m_{az}(z)$] and [$\phi_b(z)$, $m_{bx}(z)$, $m_{bz}(z)$] be the field profiles of two eigenmodes, with wavenumbers $k_{xa}$ and $k_{xb}$, of the same waveguide. Direct inspection of Eq.~\eqref{eq:governing-eqs-wg-mode} shows that $[\phi_b'(z),\linebreak[1] m_{bx}'(z),\linebreak[1] m_{bz}'(z)] \coloneqq [\phi_b(z), \linebreak[1] -m_{bx}(z), \linebreak[1] m_{bz}(z)]$ are the field profiles of an eigenmode with wavenumber $k_{xb}' = -k_{xb}$ of the complementary waveguide. Substitution of these field profiles to Eq.~\eqref{eq:lorentz-ortho-complementary} yields an orthogonality relation between two modes of the same waveguide:
\begin{multline}
 \label{eq:lorentz-ortho-same-unconjugated}
 \int_{-\infty}^\infty [\mu_0^{-1} (\phi_a b_{bx} + \phi_b b_{ax}) 
 \\
 +
  \I l^2 (k_{xa} + k_{xb}) ( -m_{ax} m_{bx} + m_{az} m_{bz})]\,\diff z = 0
  \\\text{if}\quad
  k_{xa} \neq k_{xb}.
\end{multline}

\subsection{Orthogonality relation between modes of a single waveguide (with conjugation)}

Comparison of Eqs.\ \eqref{eq:gauss-ll-a} and \eqref{eq:gauss-ll-b} shows that, \emph{in the absence of damping}, if ($\phi$, $\vect m$) satisfy Eqs.~\eqref{eq:gauss-ll-a}, then the complex-conjugate fields ($\phi^*$, $\vect m^*$) satisfy Eqs.~\eqref{eq:gauss-ll-b} governing the complementary system. Therefore if [$\phi_b(z)$, $m_{bx}(z)$, $m_{bz}(z)$] are the field profiles of a waveguide mode with wavenumber $k_{xb}$, then [$\phi_b^*(z)$, $m_{bx}^*(z)$, $m_{bz}^*(z)$] are the field profiles of a mode with wavenumber $-k_{xb}^*$ of the complementary waveguide. Substitution of these profiles to Eq.~\eqref{eq:lorentz-ortho-complementary} yields another orthogonality relation between two modes of the same waveguide:
\begin{multline}
 \label{eq:lorentz-ortho-same-conjugated}
 \int_{-\infty}^\infty [\mu_0^{-1} (-\phi_a b_{bx}^* + \phi_b^* b_{ax}) +
  \I l^2 (k_{xa} + k_{xb}^*) \vect m_{a} \cdot \vect m_{b}^*]\,\diff z = 0
  \\\text{if}\quad
  k_{xa} \neq k_{xb}^*.
\end{multline}
It should be stressed once again that this relation holds only when damping is neglected.

\section{Micromagnetic simulations \label{sec:micromagnetic-simulations}}

    To perform micromagnetic simulations we use the open-source mumax$^3$ environment \cite{vansteenkiste2014design}, which solves the full LL equation
        \begin{equation}
                \partial_t \mathbf{M} = -\frac{|\gamma| \mu_{0}}{1+\alpha^2} \left[ \mathbf{M} \times \mathbf{H}_\mathrm{eff} 
                + \frac{\alpha}{M_\mathrm{S}} \mathbf{M} \times (\mathbf{M} \times \mathbf{H}_\mathrm{eff} )
                \right],
                \label{eq:LL_basic}
        \end{equation}
    with the finite-difference time-domain (FDTD) method.
    
    We carry out simulations for the geometry presented in Fig.~\ref{fig:cofeb-system} modeled with two rectangular ferromagnetic slabs with dimensions and parameters described in subsection \ref{subsec:parameters}. We discretise the system into unit cells of size $2\times 100 \times 5$ nm$^3$ along the $x$, $y$, and $z$ axes, respectively. Additionally, to make the system independent of the $y$~coordinate, we impose periodic boundary conditions along the $y$~axis with $1024$ repetitions of the system image. We place the system in a spatially uniform in-plane magnetic field of value $\mu_{0}H_0=0.1$~T  aligned along the $y$~axis.  The damping coefficient~$\alpha$ from Eq.~(\ref{eq:LL_basic}) is set to $\alpha_0=0.0002$ in both magnetic domains. The length of the computational domain along the $x$~axis is $37.5$~\textmu m.  To prevent reflections from the outer boundaries of the modeled system, we introduce absorbing boundary conditions. Within each 9-\textmu m-wide absorbing boundary layer, the damping coefficient increases quadratically up to the value of $\alpha_\mathrm{edge}=0.5$ at the outer domain boundaries, $\alpha(\xi)=\alpha_0 + (\alpha_\mathrm{edge}-\alpha_0)\xi^2/L^2$,  where $\xi$ is the distance from the domain boundary and $L$ is the width of the absorbing boundary layer.
    
    We perform the simulations with a sweep over the stripe's width in the range from $0$ to $500$~nm with a step $2$~nm. The initial stage of each simulation is the relaxation, which finds a stable magnetic configuration required in each simulation's dynamic part. We excite spin waves by a steady  microwave field at frequency $f_0=17$~GHz, locally applied in an 8-nm-wide region. The antenna is placed $9.25$~\textmu m from the left boundary of the system. To achieve the steady state, we continuously excite the spin waves for $100$~ns. After this time, mumax$^3$ saves $40$~snapshots of the dynamic out-of-plane component of magnetisation of the system with the sampling interval $0.003$~ns.

    The results of micromagnetic simulations are saved in the form of a matrix that contains the $x$ component of the magnetisation as a  function of time and space, $m(t;x,y,z)$. The elements of the matrix  are real numbers. In the first step of the postprocessing, we perform the fast Fourier transform over time. This operation transforms the initial matrix from time-dependent to frequency-dependent, $\tilde{m}(f,x,y,z)$ and its elements become complex numbers. In the following calculations we only consider the slice of $\tilde{m}$ at the pumping frequency $f=f_0$. Fourier transform calculations show that only spin waves with frequency $f_0$ are excited, since the only peak in the Fourier spectrum appears at this frequency. This operation reduces the visibility of undesired numerical noise and transforms the data to a more easily interpretable form. It enables us to easily separate amplitude and phase of propagating waves, $|\tilde{m}|$ and $\mathrm{arg}(\tilde{m})$.      
    
    We calculate the transmittance in the system by dividing maximal values of the squared absolute value of magnetisation ($|\tilde{m}|^2$) obtained from simulations with the stripe and the reference simulation without the stripe. In each calculation step, values from the same interval $x\in(2.5;7.5)$~\textmu m placed in the segment 3, cf.\ Fig.~\ref{fig:cofeb-system}, are compared. We obtain the phase shift of the transmitted wave $\Delta\tsub{t}$ by comparing phases of spin waves from the reference simulation and the simulation with a stripe of a given width. The phases of these waves are calculated as a  mean value of $\mathrm{arg}(\tilde{m})$ for reference simulation results and simulation with the stripe results in the interval $x\in(2.5;7.5)$~\textmu m. Finally, we define $\Delta \varphi\tsub{t}$ as a difference between phase of the transmitted and reference waves, $\Delta \varphi\tsub{t} = \varphi\tsub{t} - \varphi\tsub{ref}$, normalized to the interval $(-180^\circ, 180^\circ]$.
    
    Calculations of the reflected wave parameters are more complicated, since the interference of incident and reflected waves is present in the segment 1.  Before calculating the reflectance, the contribution of the incident wave needs to be canceled. We achieve it by subtracting the reference simulation results from each result of the simulation with the stripe. The reflectance is then obtained by comparing the maximal absolute value of the spin wave amplitude of the reflected wave in an interval $x\in(-7.5;-2.5)$~\textmu m in the segment 1 with an analogous value from reference simulation but from another interval placed in the segment 3. The new interval $x\in(2.5;7.5)$~\textmu m is positioned at a similar distance to the stripe as the interval in the segment 1 and has the same length. This method assures that the attenuation in the layer influences both reflected and reference waves in the same magnitude in the calculations. 
    
    To calculate the phase shift of the reflected spin wave $\Delta \varphi\tsub{r}$ we use a similar approach as in Ref.~\onlinecite{Sobucki2021}. In this paper, very high reflectance is assumed, which is not the case in the current calculations. Thus, we assume different amplitudes of the incident and reflected waves. We obtain $\Delta \varphi\tsub{r}$ by fitting a formula $(I-R) + 2R\mathrm{cos}(2kx+\Delta \varphi\tsub{r})$ to the absolute value of simulation results $|\tilde m|$ in an interval $x\in(-5;-2.5)$~\textmu m in the segment 1. Here $I$ denotes the amplitude of the incident wave, and $R$ denotes the amplitude of the reflected wave. Like $\Delta \varphi\tsub{t}$, the phase shift $\Delta \varphi\tsub{r}$ is normalized to the interval $(-180^\circ, 180^\circ]$.

\input{modal-method.bbl}

\end{document}

%% file: modal-method.bbl
%